\documentclass[prb,twocolumn,floatfix,showpacs]{revtex4}
\usepackage[]{graphicx}
\usepackage[colorlinks,linkcolor={blue},citecolor={blue},urlcolor={blue}]{hyperref}

\begin{document}

\title{Growth of spatial correlations in the aging of a simple structural glass}

\author{ Azita Parsaeian and Horacio E. Castillo }
\affiliation{ Department of Physics and Astronomy, Ohio University,
  Athens, OH, 45701, USA } 
\date{\today} 

\begin{abstract}
We present a detailed numerical study of dynamical heterogeneities in
the aging regime of a simple binary Lennard-Jones glass former. For
most waiting times $t_w$ and final times $t$, both the dynamical
susceptibility $\chi_4(t,t_w)$ and the dynamical correlation length
$\xi_4(t,t_w)$ can be approximated as products of two factors: i) a
waiting time dependent scale that grows as a power of $t_w$, and ii) a
scaling function dependent on $t,t_w$ only through the value of the
intermediate scattering function $C(t,t_w)$. We find that
$\chi_4(t,t_w)$ is determined {\em only in part} by the correlation
volume. 
\end{abstract}

%
%
%

\maketitle

When a liquid's temperature is rapidly reduced, it can become
supercooled and eventually undergo a transition into a glass state. As
the transition is approached, the relaxation time and the viscosity
grow by several orders of magnitude. The glass transition is the point
at which the relaxation time of the liquid becomes longer than a fixed
laboratory timescale~\cite{Debenedetti-Stillinger_review_nature-410-259-2001}. Consequently, a system in the
glass state is out of thermodynamic equilibrium. In particular, {\em
physical aging } is observed: {\em time translation invariance (TTI)}
is broken, i.e. correlations between observables at times $t_w$ (the
{\em waiting time}) and $t$ (the {\em final time}) depend non-trivially
on both times, and not just on their difference $t-t_w$~\cite{bckm_review97}.

Several observations about the glass transition, including the rapidly
increasing relaxation timescales, the presence of non-exponential
relaxation, and the violation of Stokes-Einstein relations between
viscosity and diffusivity, have motivated the assumption that {\em
dynamical heterogeneities} are present in the supercooled
liquid~\cite{Debenedetti-Stillinger_review_nature-410-259-2001, Ediger_review00,Sillescu_review99}. {\em Dynamical
heterogeneities} are nanometer-scale regions containing molecules that
rearrange cooperatively at very different rates compared to the
bulk~\cite{Ediger_review00,Sillescu_review99}. 
More recently, 
dynamical heterogeneities have been directly observed in local probe experiments 
in supercooled liquids~\cite{Kegel-Blaaderen-science00, Weeks-Weitz},
glasses~\cite{Courtland-Weeks-jphysc03,
  Israeloff-AFM-nature2000+PRL2005, 
  wang-song-makse}, and granular systems near the jamming
transition~\cite{granular_chi4}. However, except for the pioneering
work by Parisi~\cite{parisi_jpcb99}, most simulations studying
dynamical heterogeneities in off-lattice models of structural glasses
have focused on the supercooled
regime~\cite{kob-donati-plimpton-poole-glotzer_hetdyn-lj_prl-79-2827-1997, yamamoto-onuki_supercooled-heterogeneity-rheology-diffusion_pre-58-3515-1998,
  Glotzer_etal_JChemPhys, Biroli_etal,
  whitelam-berthier-garrahan_dynamic-criticality-glass_prl-92-185705,
  Donati-Douglas-Kob-Plimpton-Poole-Glotzer_stringlike_prl-80-2338-1998,
  doliwa-heuer_hard-sphere-correlations_pre-61-6898-2000,
  flenner-szamel_anisotropic-hetdyn_jphyscm-19-205125-2007}

In this Letter we present a simulation study of dynamic spatial
correlations of a continuous-space, quasi-realistic glass model in the
aging regime, with the goal of characterizing their extent, space
dependence, and time dependence. We discuss the spatial correlations
of density fluctuations, which have been used before to characterize
dynamical heterogeneities in the supercooled
regime~\cite{Glotzer_etal_JChemPhys,Biroli_etal}.
We first present detailed results from an isotropic probe, which
allows us to focus on the strength and spatial extent of the
correlations, and later on briefly discuss the case of an anisotropic
probe, which permits studying the geometry of the correlated region.
A theoretical framework that postulates the presence of local
fluctuations in the age of the
sample~\cite{ckcc-rpg-prl02, RpG_numeric},
predicts that probability distributions of local fluctuations in the
aging regime are approximately independent of $t_w$ at fixed values of
the two-time global correlation $C_{\mbox{\scriptsize global}}(t,t_w)$. This
prediction has been confirmed in simulations of 
both spin
glasses~\cite{RpG_numeric}
and structural glasses~\cite{short_rhoC} (where $C_{\mbox{\scriptsize
global}}$ is the self part of the intermediate scattering
function). Motivated by this scaling,
we examine the possible presence of scaling behavior as a function of
$C_{\mbox{\scriptsize global}}$ and $t_w$ in the dynamic spatial
correlations. 

\begin{figure}[ht]

  \begin{center}
    \includegraphics[width=3.25in]{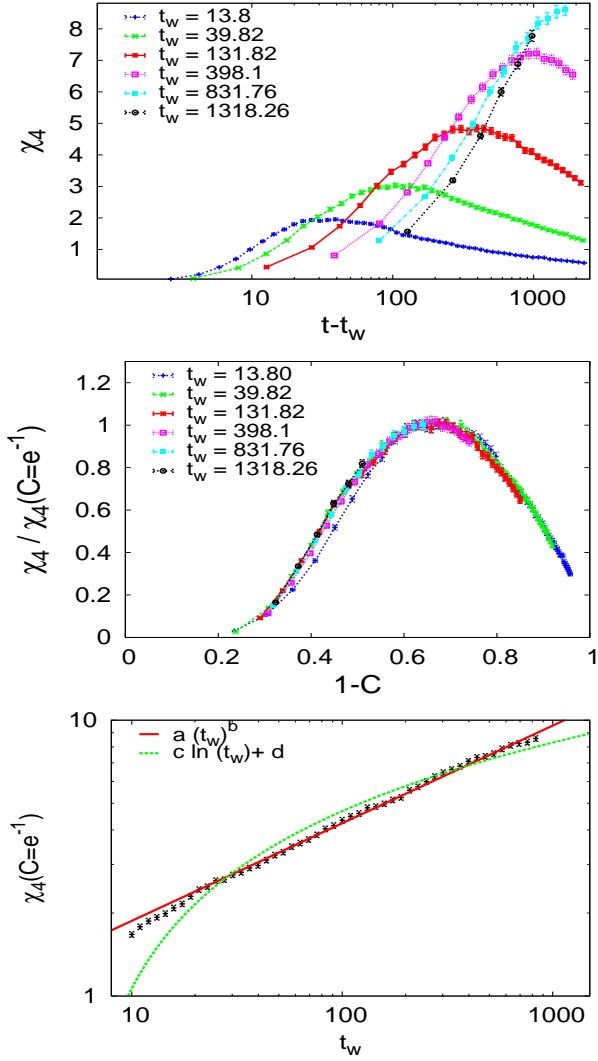}
  \end{center}
  
  \vspace{-0.5cm}
  \caption{(Color online) {\em Top panel:} Generalized density
      susceptibility $\chi_4$ as a function of the time difference
      $t-t_w$, for waiting times $t_w = 13.8, \cdots, 1318.26$. 
      {\em Middle panel:} Rescaled density
      susceptibility versus $1-C$, for the same waiting times.  {\em
      Bottom panel:} Rescaling factor of the density susceptibility,
      $\chi_4(C=1/e)$, versus the waiting time. The straight line is a
      fit to $\chi_4(C=1/e) = a {t_w}^b$, with fitting parameters
      $b=0.353 \pm 0.003$ and $a=0.83 \pm 0.02$. 
      The curved line is a
      logarithmic fit: $\xi_4(C=1/e)= c \ln (t_w) + d$, with
      parameters $c=1.57 \pm 0.04$ and $d=- 2.6 \pm 0.2$. 
      Statistical
      errors are shown by error bars in the data for all three
      panels.}
    \label{fig:chi4}

\end{figure}

\begin{figure}[ht]
  \begin{center}
    \includegraphics[width=3.25in]{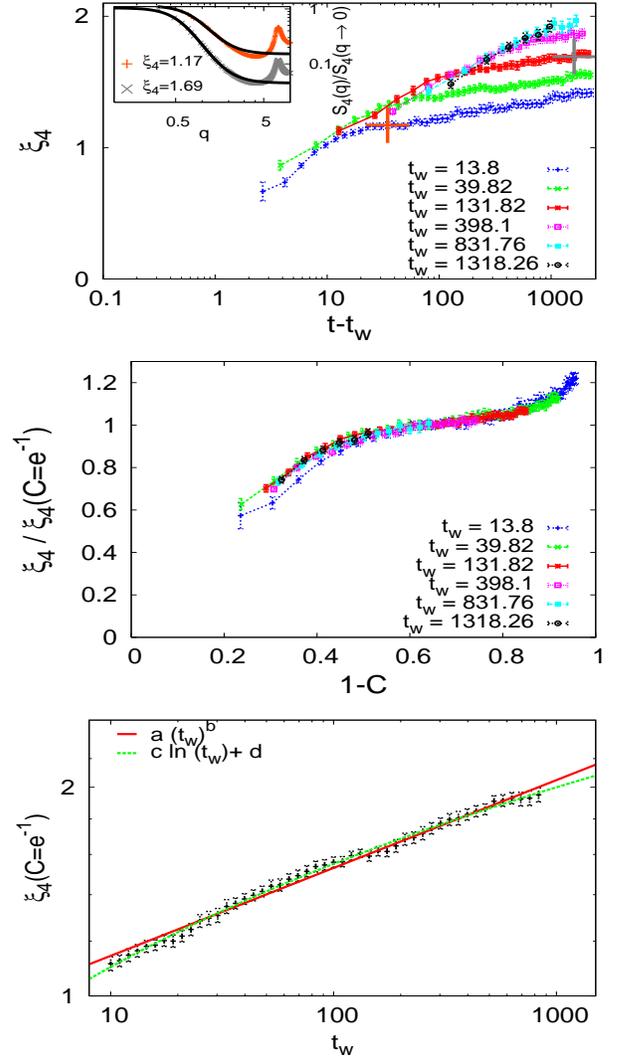}
  \end{center}
  
  \vspace{-0.5cm}
  \caption{(Color online) {\em Top panel inset:} Rescaled
    $S_{4}(q,t,t_w)$ as a function of $q$ for two $(t_w,t)$ pairs,
    with low-q fits shown (dashed lines). The correlation lengths
    extracted from the fits are indicated in the key. {\em Top panel:}
    Correlation length as a function of $t-t_w$ for various
    $t_w$'s. The $''+''$ represent the $(t-t_w, \xi_4)$ values for
    the two curves in the inset. {\em Middle panel:} Rescaled $\xi_4$
    versus $1-C$.  {\em Bottom panel:} Rescaling factor of the
    correlation length, $\xi_4(C=1/e)$, versus $t_w$. The straight
    line is a fit to $\xi_4(C=1/e) = a {(t_w)}^b$, with parameters
    $a=0.853 \pm 0.007$ and $b=0.126 \pm 0.001$.  The other line is a
    logarithmic fit: $\xi_4(C=1/e)= c \ln (t_w) + d$, with parameters
    $c=0.195 \pm 0.001$ and $d=0.651 \pm 0.008$. Statistical errors
    are shown by error bars in the data for all three panels.}
  \label{fig:xi4}
\end{figure}

\begin{figure}[ht]
  \begin{center} 
    \includegraphics[width=3.5in]{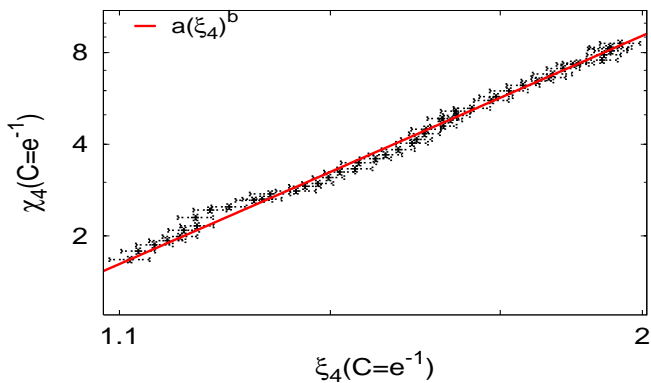}
  
    \vspace{-0.5cm}
    \caption{(Color online) 
      4-point density susceptibility rescaling
      factor versus the dynamic correlation length rescaling
      factor. The straight line corresponds to $a ( {\xi_4} ) ^{b}$, with
      $b=2.89 \pm 0.03$ and $a=1.23 \pm 0.02$. Statistical errors
      for both abscissas and ordinates are shown
      for all the data points.
    }

    \label{fig:chi4-xi4}
  \end{center} 
\end{figure}

\begin{figure*}[ht]
  \begin{center}
    \includegraphics[width=7.1in]{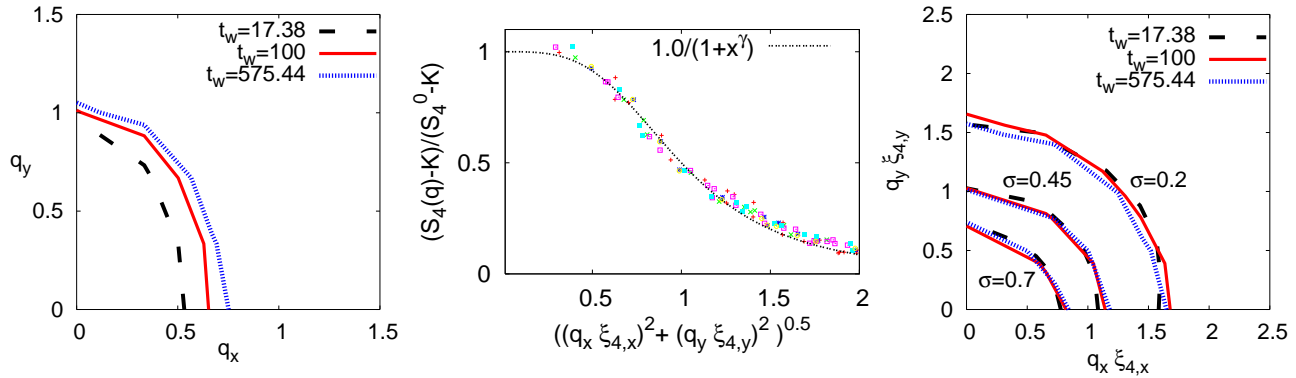}

    \vspace{-0.5cm}
    \caption{(Color online) {\em Left panel:} Contours for 
      constant
      ${\cal S}_{4}({\bf q}, {\bf k}, 
      t, t_w)$, plotted as a function of the $q_x$ and $q_y$
      coordinates, for $q_z = 0$, waiting times $t_w = 17.38, 100,
      575.44$, and constant $C_{\mbox{\scriptsize global}} = 0.3$.
      The correlations extend more in the $q_y$ direction than in the
      $q_x$ direction. {\em Middle panel:} Rescaled 4-point function
      $\sigma({\bf q}, {\bf k}, t, t_w)$ for $6$ time pairs, for $t_w
      = 17.38, 100, 575.44$, and $C_{\mbox{\scriptsize global}} = 0.3,
      0.5$. All of the rescaled data approximately collapse onto one
      curve which is fairly well fitted by the scaling form
      $1/(1+x^\gamma)$. {\em Right panel:} Contours of constant
      $\sigma({\bf q}, {\bf k}, t, t_w) =
      0.2, 0.45, 0.7$, plotted as a function of the rescaled
      wavevector $(\xi_{4,x} q_x, \xi_{4,y} q_y)$, for three
      time pairs: $t_w=17.38, 100, 575.44$ with 
      $C_{\mbox{\scriptsize global}} = 0.3$. }
    \label{fig:cdxi}
  \end{center} 
\end{figure*}

In order to probe the spatial density correlations, we use the
$4$-point ($2$-time, $2$-position) correlation
function~\cite{Glotzer_etal_JChemPhys}:
\begin{eqnarray}
  \lefteqn{ g_4({\bf r},t,t_w) =
  \frac{1}{ N \rho} \Big \langle \sum_{ik}
  \delta({ \bf r}-{\bf r}_k(t_w)+{ \bf r}_i(t_w)) }
  \\
  &&
  \times w(|{ \bf r}_i(t)-{\bf r}_i(t_w)|) 
  w(|{\bf r}_k(t)-{\bf r}_k(t_w)|)\Big \rangle 
  - \Big \langle \frac{Q(t, t_w)}{N} \Big \rangle ^2; \nonumber
\end{eqnarray}
where $ Q(t, t_w) \equiv \sum_{i=1}^N w(|{\bf r_i}(t_w)-{ \bf
r_i}(t)|)$ and $ w(|{\bf r_1}-{\bf r_2}|)$ is an overlap function
which is unity for $|{\bf r_1}-{\bf r_2}|\leq a $ and zero
otherwise. ($a \equiv 0.3\,\sigma_{AA}$ is an upper
bound for the typical amplitude of the vibrational motion~\cite{Glotzer_etal_JChemPhys}.)
By integrating $g_4({\bf r},t,t_w)$ over space, the generalized $4$-point density susceptibility
$\chi_4 \equiv \int {d^3{\bf r}} \; g_4({\bf r}, t, t_w) $~\cite{Glotzer_etal_JChemPhys} is obtained.
The self part of the intermediate scattering function $
C=C_{\mbox{\scriptsize global}}(t, t_w)=\frac{1}{N}\sum_{j=1}^{N}\exp(i{\bf
q}.({\bf r_j}(t)-{\bf r_j}(t_w))),
$
can be used as a measure of the correlation between the configurations
of the system at times $t_w$ and $t$: %
it is unity when $t=t_w$
and decays to zero when $t-t_w \gg
t_w$~\cite{Debenedetti-Stillinger_review_nature-410-259-2001, bckm_review97, 
Kob-Barrat-aging-prl97}.

We performed 4000 independent molecular dynamics runs for the binary
Lennard-Jones (LJ) system of Ref.~\cite{Kob-Barrat-aging-prl97}, which
has a mode coupling critical temperature $T_c =0.435$.  A system of
8000 particles was equilibrated at a temperature $T_0 = 5.0$, then
instantly quenched to $T = 0.4$, and then allowed to evolve at that
temperature for $2.5 \times 10^3$ Lennard-Jones time units. The origin
of times was taken at the instant of the quench. After the quench,
data were taken at times in a geometric sequence $t_n = t_0 r^n$,
(with $n=0,\ldots,61$, $t_0 =10$, $r=10^{1/25} \approx 1.09$).

We first discuss the 4-point density susceptibility $\chi_4$.  In the
top panel of Fig.~\ref{fig:chi4} we see that for fixed waiting time
$t_w$, as the time difference $t-t_w$ increases, the spatial
correlation increases till it reaches a maximum, and then
decreases. The value of the correlation $\chi_4$ at the peak and the
time difference $t-t_w$ at which the peak occurs both grow as a
function of $t_w$. Analogous behaviors are observed in numerical
simulations of supercooled liquids~(see, e.g.,
Ref.~\cite{Glotzer_etal_JChemPhys,
Biroli_etal}), as the temperature is decreased; and in experiments in
2D granular systems~\cite{granular_chi4}, as the area
fraction is increased.
As we anticipated above, we test for a possible scaling behavior with
$C = C_{\mbox{\scriptsize global}}$ by plotting, in the middle panel
of Fig.~\ref{fig:chi4}, the rescaled quantity $\chi_4/\chi_4(C=1/e)$
as a function of $1-C$.  We observe that all curves approximately
collapse onto a single master curve. Thus we can say that $\chi_4(t,
t_w) \approx \chi_4^\circ(t_w)\phi(C(t,t_w))$, where
$\chi_4^\circ(t_w) \equiv \chi_4(t,t_w)|_{C(t,t_w)=1/e}$; a rescaling
factor that depends only on $t_w$, and $\phi(C(t,t_w))$ is a scaling
function which depends on times only through the value of the
intermediate scattering function $C(t, t_w)$. As we see in the bottom
panel of Fig.~\ref{fig:chi4}, the rescaling factor as a function of
$t_w$ can be fitted with the power law form $\chi_4^\circ(t_w)=a
{t_w}^{b}$, with the power $b=0.353 \pm 0.003$~\cite{exponent-tau}. 
As the figure shows, this fit is much better than a fit to a logarithmic
form $c \ln (t_w) + d$.

We now extract a dynamic correlation length $\xi_4(t, t_w)$ from our
data.  By Fourier transforming the correlation function $g_4({\bf r},
t, t_w)$ we obtain the 4-point dynamic structure factor $S_4({\bf
q},t,t_w)$~\cite{Glotzer_etal_JChemPhys, Biroli_etal,
  Donati-Douglas-Kob-Plimpton-Poole-Glotzer_stringlike_prl-80-2338-1998}. 
We fit its small $q$ behavior with an empirical scaling
form~\cite{Biroli_etal}: $ S_4({\bf q})= \left( S^{0}_{4} - K \right)
f(\xi_4 |{\bf q}|) + K$, with $f(x) \equiv 1/(1+x^{\gamma})$.
Choosing $\gamma
= 2$ would give an Ornstein-Zernicke (OZ) form, but there is
theoretical
evidence~\cite{Biroli-Reichman-etal_inhomogeneous-mct-prl2006} that OZ
may not be a good description of $ S_4({\bf q})$, and in fact $\gamma =
2.9 \pm 0.1$ provides substantially better fits to our
data~\cite{Biroli-Reichman-etal_inhomogeneous-mct-prl2006}. Since the
${\bf q}$-dependent part of $S_4({\bf q})$ scales with $\xi_4 |{\bf
  q}|$, and its value is reduced from its maximum at $|{\bf q}|=0$ to
half of its maximum at $|{\bf q}|=1/\xi_4$, we interpret $\xi_4$ as a
correlation length.
The fits are performed in the range $0.4 \le q \le 1.9$, for 175
curves corresponding to all the time pairs shown in
Fig.~\ref{fig:xi4}. The fitting parameters $S^{0}_{4}(t,t_w)$,
$\xi_4(t, t_w)$ and $K(t,t_w)$ are determined independently for each
time pair $(t, t_w)$. The scaling function $f(x)$ is {\em not\/} allowed
to change with $(t, t_w)$: the $(t, t_w)$-independent empirical parameter
$\gamma$ is determined by minimizing the average square error over the
whole set of fits. The maximum difference between the data and the
corresponding fits never exceeds $7.5 \%$~\cite{long_corr}.
Two typical fits 
are shown in the inset of the top panel of Fig.~\ref{fig:xi4}.

The top panel of Fig.~\ref{fig:xi4} shows the correlation length
$\xi_4$ as a function of the time difference $t-t_w$ for
$t_w=13.8,...,1318.26$. We see two regimes in this plot: i) an early
aging regime~\cite{always_aging} ($t-t_w \alt t_w$), where
$\xi_4$ is approximately time translation invariant (i.e. it depends
on $t-t_w$ but {\em not} on $t_w$), and ii) a late aging regime
($t-t_w \agt t_w$), where $\xi_4$ generally grows with increasing $t_w$
at fixed $t-t_w$.  This increase of $\xi_4$ with $t_w$ is analogous to
its increase in supercooled liquids as the
temperature is lowered~\cite{Glotzer_etal_JChemPhys,
  berthier-biroli-bouchaud-cipelletti-elmasri-lhote-ladieu-pierno_science-310-1797-2005}.
                                                    
We find that $\xi_4$ always grows as a function of $t-t_w$, at fixed
waiting time $t_w$.
Simulations in the supercooled regime show divergent results on this
issue: in Ref.~\cite{Glotzer_etal_JChemPhys} the correlation length
increases initially, reaches a maximum and goes to zero for long time
differences, while in Refs.~\cite{Biroli_etal,
doliwa-heuer_hard-sphere-correlations_pre-61-6898-2000} the
correlation length always grows with time difference. Similar to
$\chi_4$, the correlation length $\xi_4$ also displays an approximate
scaling behavior with $C_{\mbox{\scriptsize global}}$: in the middle
panel of Fig.~\ref{fig:xi4}, we plot $\xi_4/\xi_4(C=1/e)$ as a
function of $1-C$.
The curves
collapse onto a single master curve and we can conclude that $\xi_4(t,
t_w) \approx \xi_4^\circ(t_w)\varphi(C(t,t_w))$, where
$\xi_4^\circ(t_w) \equiv \xi_4(t,t_w)|_{C(t,t_w) = 1/e}$.
In the bottom panel of Fig.~\ref{fig:xi4}, we observe that the
dependence of the scaling factor $\xi_4^\circ(t_w)$ on $t_w$ can be
fit roughly equally well by a power law form $\xi_4^\circ(t_w)=a
{(t_w)}^b $ where $b=0.126 \pm 0.001$~\cite{parisi_jpcb99}; or by a
logarithmic form $\xi_4^\circ(t_w)=c \ln(t_w) + d$. However the power
law form seems preferable because it is consistent with the physically
motivated scaling $\chi_4^\circ \sim (\xi_4^\circ)^{b}$ found below.

To test whether the value of $\chi_4(t, t_w)$ is controlled by the
correlation volume $(\xi_4(t, t_w))^3$, in Fig.~\ref{fig:chi4-xi4} we
plot the susceptibility scaling factor $\chi_4^\circ(t_w)$ as a
function of the correlation length scaling factor $\xi_4^\circ(t_w)$,
and find that the data are indeed well represented by a power law form
with a power close to $3$ : ${\chi_4}^{\circ} \approx
({\xi_4}^{\circ})^{b}$ with $b=2.89 \pm 0.03$ (see also
Ref.~\cite{parisi_jpcb99}). 
However, looking at the data in more
detail reveals an additional variation of $\chi_4(t, t_w)$ that cannot
be due to the correlation volume: if we compare the two plots of
$\chi_4(t, t_w)$ and $\xi_4(t, t_w)$ versus the global correlation
$C(t, t_w)$ at constant $t_w$ (the middle panels of
Figs.~\ref{fig:chi4}~and~\ref{fig:xi4}), we find that as $C(t, t_w)
\to 0$, $\chi_4(t, t_w)$ almost reaches zero, but $\xi_4(t, t_w)$
increases. Clearly, the evolution of $\chi_4$ at constant $t_w$ is
controlled not only by the correlation length but also by the
amplitude of the correlations.

We now briefly turn our attention to the anisotropic spatial structure
of the
correlations~\cite{Donati-Douglas-Kob-Plimpton-Poole-Glotzer_stringlike_prl-80-2338-1998,
doliwa-heuer_hard-sphere-correlations_pre-61-6898-2000,
flenner-szamel_anisotropic-hetdyn_jphyscm-19-205125-2007}. We define a new spatial correlator ${\cal G}_{4}({\bf
r}, {\bf k}, t, t_w)$ by replacing, in the definition of $g_4({\bf r},
t, t_w)$, all 
occurrences of $w(|{\bf r}(t) - {\bf r}(t_w)|)$
with the anisotropic function $\cos({\bf k} \cdot ({\bf r}(t) - {\bf
r}(t_w)))$~\cite{short_rhoC}.
We choose ${\bf k} = (6.68, 0.00, 2.67)$,
with $|{\bf k}| = 7.2$, i.e. at the peak of the static structure
function $S(k)$. The Fourier transform of ${\cal G}_{4}({\bf
r}, {\bf k}, t, t_w)$ with respect to ${\bf r}$ is the 4-point
anisotropic dynamic structure factor ${\cal S}_{4}({\bf q}, {\bf k}, t,
t_w)$~\cite{flenner-szamel_anisotropic-hetdyn_jphyscm-19-205125-2007}. As
we see in the left panel of Fig.~\ref{fig:cdxi}, this 4-point function
is indeed anisotropic: the contours of
constant ${\cal S}_{4}$ extend further in the $q_y$ direction than in
the $q_x$ direction.
We fitted ${\cal S}_{4}$ with an empirical form which generalizes the
isotropic one: $ {\cal S}_4 = \left( {\cal S}^{0}_{4} - K \right) f
\left([(q_x \xi_{4,x})^2+(q_y \xi_{4,y})^2 ]^{1/2}\right) + K$, with
$f(x) = 1/(1+x^{\gamma})$.  Here ${\cal S}^{0}_{4}, K, \xi_{4,x}$, and
$\xi_{4,y}$ are $(t,t_w)$-dependent fitting parameters, and
$\gamma=3.4 \pm 0.1$ (independent of $(t,t_w)$) is determined by a
global fit to all the data shown in Fig.~\ref{fig:cdxi}. The middle
panel of Fig.~\ref{fig:cdxi} shows that the rescaled quantity
$\sigma({\bf q}, {\bf k}, t, t_w) \equiv ( {\cal S}_{4}({\bf q}, {\bf
  k}, t, t_w) - K(t, t_w) )/( {\cal S}^{0}_{4}(t, t_w) - K(t,t_w) )$,
computed for six time pairs, collapses onto one curve that agrees well
with $f(x)$.  The right panel of Fig.~\ref{fig:cdxi} shows a contour
plot of $\sigma({\bf q}, {\bf k}, t, t_w)$. Here the contours for
various time pairs collapse, as long as $\sigma$ is kept constant and
rescaled wavevector components $\xi_{4,x} q_x$ and $\xi_{4,y} q_y$ are
used on the axes.
Since $k_y = 0$, and $q_z = 0$, ${\cal S}_{4}({\bf q}, {\bf k}, t,
t_w)$ is probing correlations between displacements in the $x$
direction.  We find that $\xi_{4,x}(t,t_w) > \xi_{4,y}(t,t_w)$,
i.e. the correlation extends further along the direction of the probed
displacements than along the perpendicular direction, and we interpret
this as a manifestation of cooperative ``string-like'' motion, like in
the supercooled
regime~\cite{Donati-Douglas-Kob-Plimpton-Poole-Glotzer_stringlike_prl-80-2338-1998,
  doliwa-heuer_hard-sphere-correlations_pre-61-6898-2000,
  flenner-szamel_anisotropic-hetdyn_jphyscm-19-205125-2007}.

In summary, we have studied the aging regime of a simple glass-forming
model and found evidence of scaling for the dynamical susceptibility
$\chi_4(t,t_w)$ and the dynamical correlation length $\xi_4(t,t_w)$ as
products of a power of $t_w$ times a scaling function that depends on
times only through $C_{\mbox{\scriptsize
    global}}(t,t_w)$. Additionally, we have found evidence of
time-independent scaling functions that describe the ${\bf q}$
dependence of both the isotropic and the anisotropic dynamic structure
factors. In the anisotropic case, we have found evidence for
``string-like motion'', similar to the one observed in supercooled
liquids.

H.~E.~C. especially thanks C.~Chamon and L.~Cugliandolo for very
enlightening discussions over the years, and J.~P.~Bouchaud,
S.~Glotzer, N.~Israeloff, M.~Kennett, D.~Reichman, and E.~Weeks for
suggestions and discussions. 
This work was supported in
part by DOE under grant DE-FG02-06ER46300, by NSF under grant PHY99-07949,
and by Ohio University. 
Numerical simulations
were carried out at the Ohio Supercomputing Center and at the Boston
University SCV.  H.~E.~C. acknowledges the hospitality of the Aspen
Center for Physics.


\begin{thebibliography}{99}
\bibitem{exponent-tau} 
  In our data~\cite{long_corr} $\tau_{\alpha}(t_w) \sim t_w^{0.9}$,
  which leads to 
  $\chi_4 \sim (\tau(t_w))^{0.4}$. For a supercooled LJ system $\chi_4
  \sim (\tau(T))^{0.4}$ was found~\cite{whitelam-berthier-garrahan_dynamic-criticality-glass_prl-92-185705}.
\bibitem{always_aging} 
  For small enough $t-t_w$, $C_{\mbox{\scriptsize global}}(t,t_w)$
  should be TTI (i.e. no aging), but in our data time separations are
  too large to observe this regime.
\end{thebibliography}
\end{document}